\documentclass[
    ,final            
  ]
  {aipproc}

\layoutstyle{6x9}

\begin{document}

\title{The Arecibo Legacy Fast ALFA (ALFALFA) Extragalactic HI Survey}


\author{Martha P. Haynes and the ALFALFA Team}{
  address={530 Space Sciences Building, Cornell University, Ithaca NY 14853}
}

\begin{abstract}
The Arecibo Legacy Fast ALFA (ALFALFA) survey is a program aimed at
obtaining a census of HI-bearing objects over a cosmologically significant
volume of the local universe. When complete in $\sim$3--4 years, it will
cover 7000 square degrees of high latitude sky using the 305~m telescope
and the seven-beam Arecibo L-band feed array (ALFA). As of May 1, 2008, almost 60\%
of the required observations are complete and a catalog exists in preliminary
form for 25\% of the final sky area.
ALFALFA is detecting about twice as many HI sources as predicted based
on previously published HI mass functions and should deliver a final
catalog of $> 25000$ extragalactic HI sources. ALFALFA will detect hundreds of galaxies
with HI masses less than $10^{7.5} M_\odot$ and similarly large numbers 
greater than $10^{10.3} M_\odot$. Its centroiding accuracy
allows for the immediate identification of highly probably optical counterparts
to each HI detection.
Fewer than 3\% of all extragalactic HI sources, and $<$ 1\% of 
ones with $M_{HI} > 10^{9.5} M_\odot$ cannot be identified with a
stellar counterpart. The hundreds of HI sources with observed line widths of 
$20-30$ km~s$^{-1}$ include a population of optically faint dwarf galaxies.
The objects with highest HI masses exhibit a range
of morphologies, optical colors and surface brightnesses, but most appear
to be massive disk systems. The latter represent the population likely
to dominate future studies of HI at high redshift.
\end{abstract}

\maketitle


\section{The ALFALFA Survey Program}

ALFALFA, the Arecibo Legacy Fast ALFA Survey, is a two-pass spectral line survey
to cover 7000 deg$^{2}$ of high galactic latitude sky\cite{Gio05a} with $\sim$eight
times the sensitivity, four times the angular resolution, three times the spectral resolution,
and 1.6 times the total bandwidth of the HI Parkes All-Sky Survey (HIPASS)\cite{Barnes01}.
The ALFALFA survey strategy has been designed specifically to exploit Arecibo's superior
sensitivity, angular resolution and digital technology to conduct a
census of the local HI universe over a cosmologically significant volume\cite{Gio08}\cite{Hay08}.
The effective integration time is $\sim$40 sec
per beam area, yielding approximately 2.2 mJy at 10 km~s$^{-1}$ resolution (after Hanning smoothing). 
The survey is intended to map, with complete 2-pass coverage, the region from 0$^\circ$ to +36$^\circ$
in declination and from $22^h <$R.A.$< 3^h$  and
$7^h30^m <$R.A.$< 16^h30^m$. The fixed azimuth, ``minimum intrusion''
observing technique\cite{Gio05b}\cite{Gio07} delivers extremely high data quality 
and observing efficiency (99\% ``open shutter'' time). Because of its
wide areal coverage and photometric accuracy, ALFALFA is providing a legacy dataset for
the astronomical community at large, serving as the basis for numerous studies of the
local extragalactic Universe. The survey was initiated in February 2005; as of May 1, 2008,
$\sim 60$\% of the survey observations have been completed.

The ALFALFA team is an open 
collaboration of more than 60 researchers from 34 institutions in 13 countries.
Anyone with an interest in the science that can be done with the ALFALFA dataset 
and the willingness to contribute to the collective effort is welcome to join.
Guidelines for joining ALFALFA can be found at \url{http://egg.astro.cornell.edu/alfalfa/joining.php}.
The ALFALFA survey is also serving as the backbone for student research projects at
both the graduate and undergraduate level.
One Ph.D. thesis\cite{AS07t} based on ALFALFA is already complete; currently ten graduate
students from six different institutions are working on Ph.D. dissertations centered
on ALFALFA research. Current ALFALFA team projects are summarized at
\url{http://egg.astro.cornell.edu/alfalfa/projects/teamprojects.php}. The undergraduate
ALFALFA program supports the participation of faculty and students at 14 institutions
engaged in research, observing and educational exchange. An undergraduate workshop
\url{http://egg.astro.cornell.edu/alfalfa/ugradteam/ugradj08.php} is held at Arecibo
each year. Eight ALFALFA senior theses have been undertaken so far.

Data processing for ALFALFA
makes us of the IDL-based ALFALFA pipeline developed at Cornell and
exported successfully to more than 16 institutions running both 
Linux and MacOS. Heavy use is made of Virtual Observatory
protocols and web
services for real-time cross-correlation with public multiwavelength 
databases. The identification of HI sources in the gridded data is performed 
using a Fourier domain matched filter signal extraction technique
\cite{AS07}.  Simultaneously, the most probable optical counterpart, where
such exists, is identified using public OIR survey datasets.
In addition to the most reliable, high S/N detections,
sources of lower S/N but which coincide in both position and redshift with
known optical galaxies are also included, but flagged as such, in the catalog of detections. 
Data catalogs and products are available 
at \url{http://arecibo.tc.cornell.edu/hiarchive/alfalfa/}.

Two catalogs of HI sources extracted from 3-D spectral data cubes have been
published\cite{Gio07}\cite{AS08} and a third has been submitted\cite{Kent08}.
Several additional publications should be submitted this summer, including
a completed catalog of the $\sim$1600 square degrees in the region 
$7^h30^m <$R.A.$< 16^h30^m$, $+04^\circ <$ Dec $< +16^\circ$. In the latter
area, ALFALFA detects $> 6200$ high quality sources versus 290 for HIPASS.

\section{ALFALFA: Early Science}

Although still in the early stages, ALFALFA is already delivering on
its promised scientific harvest. Here are some of the most
interesting results to date.

{\bf HI census:} \hskip 5pt  ALFALFA is now on course to detect $>25000$ 
sources, twice as many objects as we predicted
\cite{Gio05a} based on the HI mass functions derived from previous surveys.
Fewer than 3\% of all extragalactic HI sources, and $<$ 1\% of 
ones with $M_{HI} > 10^{9.5} M_\odot$ cannot be identified with a
stellar counterpart.

{\bf A Blind HI Survey of the Virgo Region:} \hskip 5pt The initial ALFALFA results cover
the central region of the Local Supercluster, in and around the Virgo Cluster\cite{Gio07}
\cite{Kent08}. A number of
extensive ($> 250$ kpc) HI streams or complexes, suggestive of tidal or
``harassment'' debris, have been discovered \cite{Kent07}
\cite{Hay07}\cite{Koop08}, all on the outskirts of the cluster. On-going work includes
derivation of the HI mass function in Virgo; a significant absence of high HI mass
systems is apparent, caused by the well-known HI deficiency of cluster spirals\cite{Kent08t}.
Members of the ALFALFA collaboration are exploiting multiwavelength data and numerical
simulations to explore the detailed properties of the galaxy population and the mechanisms
likely at play in the Virgo region\cite{Sper07}\cite{GG08}.

{\bf Dwarf galaxies and high velocity clouds:} \hskip 5pt 
ALFALFA is specifically designed to detect very
low mass galaxies in the local universe. Besides its sensitivity advantage, its
superior spectral resolution allows detection of HI lines as narrow (FWHM) as 
20 km~sec$^{-1}$\cite{AS07t}\cite{AS07},
characteristic of low mass halos. Members of the ALFALFA team are
conducting a coordinated campaign of optical imaging and long-slit spectroscopy,
GALEX imaging and HI synthesis studies to probe the impact of local environment of the 
population of low HI mass, gas-rich dwarf galaxies
discovered by ALFALFA\cite{AS07t}. ALFALFA is cataloguing high velocity clouds associated
with the Milky Way; several such clouds close to M33 have also been discovered 
in the ALFALFA dataset \cite{Gro08}.

{\bf High HI mass galaxies} \hskip 5pt 
ALFALFA detects galaxies with HI masses as high as $10^{10.8}M_\odot$, representative
of the massive disks likely to be studied at high redshift. Many of these
are large, luminous galaxies with well-delineated spiral arms. Some have extended,
low surface brightness disks. As a class, these objects
provide a glimpse of the gas-rich component of the ``transition mass'' systems 
targeted by the GASS survey
presented by David Schiminovich and the optically-selected gas-rich galaxies detected at
Arecibo at $z \sim 0.2$ as reported
by Barbara Catinella in this volume.


\begin{theacknowledgments}
My participation in ALFALFA is supported by NSF grant AST-0607007
and by the Brinson Foundation. The Arecibo Observatory
is part of the National Astronomy and Ionosphere Center which is
operated by Cornell University under a cooperative agreement
with the National Science Foundation. The US National Virtual Observatory
is sponsored by the National Science Foundation.
I thank all of the members of the
ALFALFA team for their enthusiasm for its science and for their
huge efforts in its pursuit. 
\end{theacknowledgments}



\bibliographystyle{aipproc}   

\begin{thebibliography}{9}

\bibitem{Gio05a}
Giovanelli, R.,  {\it et al.}, 2005, \emph{AJ} 130, 2589.

\bibitem{Barnes01}
Barnes, D.G, {\it et al.}, 2001, \emph{MNRAS}, 322, 486.

\bibitem{Gio08}
Giovanelli, R. 2008, \emph{Il Nuovo Cimento} (in press)

\bibitem{Hay08}
Haynes, M.P. 2008, \emph{Il Nuovo Cimento} (in press)

\bibitem{Gio05b}
Giovanelli, R.,  {\it et al.}, 2005, \emph{AJ} 130, 2613.

\bibitem{Gio07}
Giovanelli, R.,  {\it et al.}, 2007, \emph{AJ} 133, 2569.

\bibitem{AS07t}
Saintonge, A. 2007, Ph.D. thesis, Cornell University.

\bibitem{AS07}
Saintonge, A. 2007, \emph{AJ} 133, 2087.

\bibitem{AS08} 
Saintonge, A., {\it et al.} 2008, \emph{AJ} 135, 588.

\bibitem{Kent08}
Kent, B.R., {\it et al.} 2008,  \emph{AJ} (submitted).

\bibitem{Kent07}
Kent, B.R., {\it et al.} 2007, \emph{Ap.J.(Lett.)} 665, L15-18.

\bibitem{Hay07}
Haynes, M.P., Giovanelli, R. \& Kent, B.R. 2007 \emph{Ap.J.(Lett.)} 665, L19-22.

\bibitem{Koop08}
Koopmann, R.A., {\it et al.} 2008, \emph{Ap.J.(Lett.)} (submitted).

\bibitem{Kent08t}
Kent, B.R., 2008, Ph.D. thesis, Cornell University (in preparation).
 
\bibitem{Sper07}
 di Serego Alighieri, S., 
{\it et al.}, 2007, \emph{AAp} 474, 851.

\bibitem{GG08}
Gavazzi, G.,  {\it et al.}, 2008, \emph{AAp} 482, 43

\bibitem{Gro08}
Grossi, M. {\it et al.}, 2008, \emph{AAp} (in preparation)

\end{thebibliography}


\end{document}